\documentclass[9pt]{article}
\usepackage{amsmath}
\usepackage{graphicx}
\usepackage{cite}  % For compact citation ranges like [1-3]
\usepackage{hyperref}  % For hyperlinked references

\usepackage[utf8]{inputenc}
\usepackage{longtable}
\usepackage{booktabs}
\usepackage{ragged2e}
\usepackage{tablefootnote}
\usepackage{array}
\usepackage{multirow}
\usepackage{tabularx}
\usepackage{xcolor}
\usepackage{url}
\usepackage{makecell}

\newcommand{\spconfBibFontSize}{6.15pt}
\usepackage{paper}

\newcolumntype{R}[1]{>{\raggedleft\arraybackslash}p{#1}}
\newcolumntype{L}[1]{>{\raggedright\arraybackslash}p{#1}}

\usepackage{microtype}            % better line breaks and character protrusion
\usepackage[font=small,labelfont=bf]{caption}
\newlength{\topadj}
\usepackage{enumitem}
\setlist{nosep,leftmargin=*}

\AtBeginDocument{\setlength{\tabcolsep}{3.5pt}}
\renewcommand{\arraystretch}{0.95}

\title{Low-Resource Audio Codec (LRAC): 2025 Challenge Description}

\begin{NoHyper}
\name{\begin{tabular}{c}Kamil Wojcicki$^{1}$, Yusuf Ziya Isik$^{1}$, Laura Lechler$^{1}$, Mansur Yesilbursa$^{1}$, Ivana Bali\'{c}$^{1}$,\\
Wolfgang Mack$^{1}$, Rafał Łaganowski$^{1}$, Guoqing Zhang$^{1}$, Yossi Adi$^{2}$, Minje Kim$^{3}$, Shinji Watanabe$^{4}$\end{tabular}}
\address{%
$^1$Collaboration AI, Cisco Systems, Inc.\\
$^2$Hebrew University of Jerusalem, Israel\\
$^3$University of Illinois Urbana-Champaign, United States\\
$^4$Carnegie Mellon University, United States} 
\end{NoHyper}

\begin{document}
\ninept
\maketitle
\begin{abstract}
While recent neural audio codecs deliver superior speech quality at ultralow bitrates over traditional methods, their practical adoption is hindered by obstacles related to low-resource operation and robustness to acoustic distortions. Edge deployment scenarios demand codecs that operate under stringent compute constraints while maintaining low latency and bitrate. The presence of background noise and reverberation further necessitates designs that are resilient to such degradations. The performance of neural codecs under these constraints and their integration with speech enhancement remain largely unaddressed. To catalyze progress in this area, we introduce the 2025 Low-Resource Audio Codec Challenge, which targets the development of neural and hybrid codecs for resource-constrained applications. Participants are supported with a standardized training dataset, two baseline systems, and a comprehensive evaluation framework. The challenge is expected to yield valuable insights applicable to both codec design and related downstream audio tasks.
\end{abstract}


\section{Introduction}

Speech coding remains a cornerstone of modern communication, enabling high-quality audio transmission under bandwidth-limited conditions. Although traditional codecs based on signal processing and perceptual models, such as Opus \cite{opus} or AMR-WB \cite{bessette2003adaptive}, achieve low-compute and low-latency operation, their performance degrades at low bitrates \cite{Zeghidour2022}.
 
Recent advancements have incorporated neural components into conventional pipelines via noise-shaping networks~\cite{buthe2023lace,Buthe2024}, neural enhancement to suppress coding artifacts \cite{Zhao2019, Korse2020, Biswas2020, Korse2022}, and neural vocoders \cite{Kleijn2018, Klejsa2019, Garbacea2019,  Valin2019a,9632750}. These \emph{hybrid} approaches can improve perceptual quality, though they may introduce some additional complexity and remain constrained by traditional coding structures, which can limit their scalability at ultralow bitrates \cite{kim2025neural}.

Fully neural end-to-end codecs, such as SoundStream~\cite{Zeghidour2022}, EnCodec~\cite{Defossez2022}, and Descript audio codec~\cite{kumar2023DAC}, demonstrate that learned analysis-synthesis pipelines can achieve high quality at low bitrates.
Subsequent systems, such as BigCodec \cite{xin2024bigcodec}, extend this paradigm and show further quality gains at extremely low bitrates~\cite{Ai2024,Wu2023audiodec,xin2024bigcodec,kumar2023DAC,parker2024stablecodec,siuzdak2024snac,Liu2024semanticodec,defossez2024moshi}. However, such models often come with increased computational and memory demands, which challenge their deployment on power-, memory- and latency-limited edge devices.

Meanwhile, recent work has also focused on \emph{neural audio tokenizers} \cite{mousavi2025discreteaudiotokenssurvey, ye2025codecdoesmatter, Liu2024semanticodec} capable of compressing speech and/or audio into discrete representations for multimodal foundation models. 
While such approaches enable a variety of speech-driven downstream tasks, they are rarely designed for real-time operation in low-resource environments and often assume access to cloud-scale computational resources.

Although this emerging research shows substantial promise, there remains a lack of benchmarks designed for the requirements of real‑time codec deployment in low‑resource settings. The \textbf{2025 Low-Resource Audio Codec (LRAC) Challenge} directly targets this gap by benchmarking codecs under \emph{joint resource constraints}---including computational complexity, bitrate, and latency---and by conducting a comprehensive evaluation aligned with real-world acoustic scenarios.
This focus distinguishes LRAC from the Codec-SUPERB Challenge~\cite{codecsuperb}, which emphasizes perceptual quality and information preservation without explicit compute or latency restrictions.

In summary, the \textbf{novel} aspects of the \textbf{LRAC Challenge} are:

\begin{enumerate}
    \item \textbf{Joint resource constraints:} 
    Includes codec development tasks under concurrent constraints on compute, latency, and bitrate.
    \item \textbf{Robustness to real-world acoustic conditions:} Codecs are evaluated for their robustness to everyday noise and reverberation.
    \item \textbf{Integrated speech enhancement and coding:} A dedicated track for codec systems that additionally perform denoising and dereverberation---a direction largely underexplored in prior work.
    \item \textbf{Listener-based benchmarking}: Use of large-scale state-of-the-art crowdsourced test battery for assessing speech quality, intelligibility, and robustness to noise and reverberation.
\end{enumerate}

This challenge series aims to foster vital research in low-resource audio coding. The first edition focuses on a subset of core speech coding advancements; complementary aspects such as expansion to general audio coding, network resilience, memory footprint, or wider system integration considerations are left for future editions. We are nevertheless confident that the 2025 LRAC Challenge will make a meaningful contribution to the advancement of practical speech communication technologies. The insights gained are also expected to be transferable to downstream applications, such as spoken dialogue systems and multi-modal language models.

\section{Challenge Description}
\vspace{2.5mm}
\label{sec:challenge-description}

\subsection{Challenge Tracks}
\vspace{0.5mm}
\label{sec:tracks}

The challenge is composed of the following two tracks.

\textbf{Track 1} focuses on developing speech codecs that prioritize transparency, aiming to minimize perceptual degradation while meeting strict constraints on computational complexity, latency, and bitrate as specified in the challenge rules \cite{lrac2025rules}. The objective is to achieve high-quality speech reproduction under relatively clean conditions. That is, the input signals are contaminated by background noise and reverberation only to a mild degree, based on the assumption that stronger distortions will have been sufficiently mitigated by prior enhancement processing in the audio pipeline, e.g., beamforming and/or single-channel noise reduction. This track acts as a benchmark for codecs that do not perform active speech enhancement, while they still need to be robust to mild noise and reverberation conditions. 

\textbf{Track 2} encourages codecs to integrate speech enhancement capabilities and compression more actively. The aim is to enable transmission of intelligible and natural-sounding speech, including under challenging acoustic environments, with an expanded computational complexity and latency budget compared to Track 1. The objectives include prioritizing transparency in clean conditions, robustness to everyday distortions such as additive noise and reverberation, and effective denoising and dereverberation performance under challenging conditions. Both end-to-end neural codecs and modular pipelines (e.g., codecs with enhancement front-ends) are welcome.

\begin{table}[t]
\vspace{\topadj}
\footnotesize
\caption{Summary of bitrate, complexity and latency constraints.}
\label{table:constraints}
\centering
\renewcommand{\arraystretch}{1.1}
\begin{tabular}{L{1.7cm} L{1.65cm} L{2.1cm} L{2.3cm}@{}}
\toprule
\multicolumn{2}{c}{\textbf{Constraint}}&\textbf{Track 1} & \textbf{Track 2} \\
\midrule
\multirow{2}{1.7cm}{\textbf{Bitrate Mode}} & Ultralow & $\le$\,1 kbps & $\le$\,1 kbps\\\cline{2-4}
& Low \rule{0pt}{2.5ex} & $\le$\,6 kbps & $\le$\,6 kbps \\
\midrule
\multirow{2}{1.7cm}{\textbf{Compute Complexity}} & Receive-side & $\le$\,300 MFLOP/s & $\le$\,600 MFLOP/s \\\cline{2-4}
& Total\rule{0pt}{2.5ex} & $\le$\,700 MFLOP/s & $\le$\,2600 MFLOP/s \\
% \addlinespace[0.5em]
\midrule
\textbf{Latency} & Total & $\le$\,30 ms & $\le$\,50 ms \\
\bottomrule
\end{tabular}
\end{table}

\subsection{Challenge Rules and Guidance}
\label{sec:rules}

Submissions were required to comply with the following key rules: 
\begin{itemize}
    \item Support for mono audio input/output at 24 kHz sampling rate.
    \item Meeting bitrate, complexity, and latency constraints in Table~\ref{table:constraints}.
    \item For simplicity of evaluation, only constant bitrate systems were permitted in the first edition of the challenge. No variable or adaptive bitrate coding or entropy coding optimization were allowed. 
    \item A single system had to support \textsl{both} the ultralow bitrate (ULB) and low bitrate (LB) modes listed in Table~\ref{table:constraints}. 
    \item The same decoder had to be capable of supporting both bitrate modes, and a mixture of the modes within a single inference run.
    \item Use of \textsl{only} officially designated speech \& noise datasets for training, validation and hyperparameter tuning for a fair comparison.
    \item Techniques like domain adaptation, self-training, or test-time adaptation involving the dev/test data were not allowed.
    \item Hybrid solutions, integrating neural and traditional coding or post-processing, were permitted, provided they met overall constraints.
    \item Participants had to submit detailed system reports.
    \item Participants were not required but were encouraged to submit peer-reviewed workshop papers and to consider open-sourcing their models with a research-permissive license.
\end{itemize}

The full set of the official rules for the challenge was made available on the challenge website \cite{lrac2025rules}. Ensuring adherence to the rules was based on the honor system and was the responsibility of each participating team.

Participants were advised that both latency and computational complexity should be calculated using analytical methods based on the model’s architecture in order to achieve an accurate and hardware-independent comparison \cite{lrac2025faqlatencycompute}. Latency was defined as a theoretical estimate only, a combination of algorithmic and buffering latencies, and excluding actual inference time and runtime variability. Computational complexity was measured as the total number of multiply-accumulate (MAC) operations, covering matrix-matrix and matrix-vector multiplications and summations, and converted from MMAC/s to MFLOP/s by multiplying by two. As noted in the Challenge FAQ \cite{lrac2025faqlatencycompute}, the contribution of nonlinear activations could be ignored since their cost is highly hardware- and implementation-dependent, and they can often be approximated by cheaper functions without significant performance impact. No specific FLOP-counting tool was mandated; instead, formula-based calculations were recommended as the primary approach, with FLOP-counting tools used only as optional guides due to potential inaccuracies. To further assist participants, the organizers provided a spreadsheet in the baseline repository with worked examples of latency and complexity calculations using layer design formulas for the Track 1 and Track 2 baseline systems \cite{lrac2025baseline}. Detailed latency calculation guidelines were also provided \cite{lrac2025latencycalcguidelines}.

\subsection{Participation and Timeline}

The challenge timeline is outlined in Table~\ref{tab:timeline}. Teams registered via a designated delegate who managed registration, correspondence, and submissions. Submissions were made through a Codabench-based \cite{codabench} leaderboard website and included system audio outputs and detailed system description reports. 

\begin{table}[bt]
  \vspace{\topadj}
  \footnotesize
  \centering
  \caption{Timeline of the 2025 LRAC Challenge.}
  \label{tab:timeline}
  \begin{tabular}{L{5cm} L{2.75cm} L{}@{}}
    \toprule
    \textbf{Phase} & \textbf{Date(s)} \\
    \midrule
    Registration deadline & Sep 9, 2025 \\
    Development (objective evaluation) & Aug 1 -- Sep 30, 2025 \\
    Test (subjective evaluation) & Oct 1, 2025 \\
    Results announced & Oct 14, 2025 \\
    Optional paper submission deadline & Oct 22, 2025 \\
    ICASSP 2026 Satellite Workshop & May 4, 2026 \\
    \bottomrule
  \end{tabular}
\end{table}

\section{Data}
\label{sec:data}

\subsection{Training Data}

\label{sec:training-data}
The challenge provided curated public datasets comprising clean speech, noise, and room impulse responses (RIRs). Use of full or filtered (curated) datasets and of data augmentations was permitted.

\begin{table}[!b]
  \vspace*{3mm}
  \centering
  \footnotesize
  \caption{Training speech data curation by dataset.}
  \label{tab:lrac-training-data}
  \begin{tabular}{l R{1.4cm} R{1.39cm} R{1.68cm}}
    \toprule
    \textbf{Dataset} & \textbf{Original (h)} & \textbf{Curated (h)} & \textbf{Retention (\%)} \\
    \midrule
    \iftrue
    GLOBE V2 \cite{wang2024globe}           & 520.9 & 186.4  & 35.8 \\
    MLS \cite{pratap2020mls} (FR, DE, ES)   & 427.8 & 275.6  & 64.4 \\
    Librivox (DNS5) \cite{dns2023}          & 313.9 & 85.3   & 27.2 \\
    LibriTTS \cite{zen2019libritts}         & 191.3 & 46.3   & 24.2 \\
    EARS \cite{richter2024ears}             & 86.8  & 86.8   & 100.0 \\
    VCTK \cite{veaux2017vctk}               & 78.8  & 22.3   & 28.3 \\
    \else
    MLS \cite{pratap2020mls} (FR, DE, ES)   & 427.8 & 275.6  & 64.4 \\
    GLOBE V2 \cite{wang2024globe}           & 520.9 & 186.4  & 35.8 \\
    EARS \cite{richter2024ears}             & 86.8  & 86.8   & 100.0 \\
    Librivox (DNS5) \cite{dns2023}          & 313.9 & 85.3   & 27.2 \\
    LibriTTS \cite{zen2019libritts}         & 191.3 & 46.3   & 24.2 \\
    VCTK \cite{veaux2017vctk}               & 78.8  & 22.3   & 28.3 \\
    \fi
    \midrule
    \textbf{Total}                          & \textbf{1641.7} & \textbf{702.7} & \textbf{42.8} \\
    \bottomrule
  \end{tabular}
\end{table}

\begin{table*}[!t]
\vspace{\topadj}
\caption{Subjective evaluation battery of the test phase for Track 1 and Track 2. ULB: ultralow bitrate, LB: low bitrate.}
\centering
\scriptsize
\renewcommand{\arraystretch}{1.3}
\begin{tabularx}{\textwidth}{l l c c X c c c}
\toprule
 & \textbf{Conditions} & \textbf{\# Files} & \textbf{Max. Bitrates} & \textbf{Goal} & \textbf{Test} & \textbf{Test Range} & \textbf{Weight (ULB / LB)} \\
\midrule

\multirow{4}{*}{\textbf{Track 1}} 
 & \textbf{1a.} Clean speech & 90 & 1, 6 & Quality on clean speech & MUSHRA--1S & [0, 100] & 20\% / 20\% \\
 & \textbf{1b.} Real-world light noise \& reverb & 90 & 1, 6 & Robustness to light noise \& reverb & DCR & [1, 5] & 20\% / 20\% \\
 & \textbf{1c.} Simultaneous talkers & 20 & 1, 6 & Preservation of multiple speakers & DCR & [1, 5] & 5\% / 5\% \\
 & \textbf{1d.} Speech intelligibility in clean & 384 & 1 & Speech intelligibility (English) & DRT & [-100, 100] & 10\% / -- \\
\midrule

\multirow{5}{*}{\textbf{Track 2}} 
 & \textbf{2a.} Clean speech & 90 & 1, 6 & Quality on clean speech & MUSHRA--1S & [0, 100] & 10\% / 15\% \\
 & \textbf{2b.} Real-world reverb & 60 & 1, 6 & Dereverberation performance & ACR & [1, 5] & 10\% / 20\% \\
 & \textbf{2c.} Real-world speech in noise & 100 & 1, 6 & Denoising performance & ACR & [1, 5] & 10\% / 20\% \\
 & \textbf{2d.} Speech intelligibility in clean & 384 & 1 & Speech intelligibility (English) & DRT & [-100, 100] & 5\% / -- \\
 & \textbf{2e.} Speech intelligibility in noise & 384 & 6 & Speech intelligibility (English) & DRT & [-100, 100] & -- / 10\% \\
\bottomrule
\end{tabularx}
\label{tab:evaluation}
\vspace{-5mm}
\end{table*}

\subsubsection{Speech data and curation}

The clean-speech training dataset selections were inspired by the 2025 URGENT Challenge~\cite{urgent2025}. Approximately $702.7$ hours clean-speech training set was curated from LibriTTS \cite{zen2019libritts}, VCTK \cite{veaux2017vctk}, EARS \cite{richter2024ears}, Librivox (DNS5) \cite{dns2023}, MLS \cite{pratap2020mls} (FR, DE, ES), and GLOBE~V2 \cite{wang2024globe} public datasets with durations listed in Table \ref{tab:lrac-training-data}. Internal audio metrics were used for curation. These included dataset-specific thresholds on signal-to-noise ratio (SNR), reverberation, bandwidth ($>$$9.5$\,kHz), sample rate ($\geq$$24$~kHz), duration ($>$$3$\,s), and saturation. We verified speaker disjointness across splits and harmonized metadata across corpora (language tags, speaker IDs, transcript---where available; and gender---where available or estimated).
In addition to quality-based filtering, speaker balancing was applied to improve the gender representation and reduce speaker dominance. A summary of post-curation durations is given in Table~\ref{tab:lrac-training-data}.

Although care was taken to provide diverse material, the curated speech set is \emph{predominantly read speech}:
LibriTTS, VCTK, Librivox (DNS5), and MLS together account for $429.5$ hours of the $702.7$ hours total ($\approx\!61\%$). Furthermore, English dominates, with some additional language coverage (FR/DE/ES) from the MLS dataset. 

Through careful curation to enhance both the quality and diversity of the training speech data, we expect the provided 702.7 hours clean-speech training corpus to be sufficient for developing low-compute neural audio codecs (NACs), consistent with recent evidence that underscores the importance of data curation in scaling the performance of speech enhancement models \cite{li2025moredatacurationmatters}.

\subsubsection{Reverb and noise data and curation}
\label{sec:reverb-noise-data-curation}

VCTK Noises \cite{Valentini-Botinhao2017-xl}, subsets of Audio Set and FreeSound used in the DNS5 Challenge \cite{dns2023}, WHAM! \cite{Wichern2019WHAM}, FSD50K\cite{fonseca2022FSD50K}, and Free Music Archive\cite{fma_dataset} were permitted for use in noise-based augmentation. 
A mixture of real and synthetic RIRs from the Motus \cite{motus} and OpenSLR 28 \cite{openslr28} datasets was suggested for reverb-based augmentation. Participants were permitted to use other RIR datasets or synthesized RIR data.

Noise datasets were curated as follows. All noise files were classified using a simplified ontology derived from the Audio Set \cite{gemmeke2017audioset}.
Noise classification was performed with CLAP~\cite{clap,taylor25_interspeech}, and files from the most frequent categories were sampled less frequently to improve the overall balance. Further, recordings with a high likelihood of containing speech were excluded, reducing the volume of the noise datasets from $518$ to $335$ hours. The interested reader is referred to Figs.~6 and 7 in \cite{lrac2025datasets} for an illustration of noise categories and the distribution of the noise data before and after curation, respectively.
The curation helped achieve a diverse noise dataset spanning major noise categories.

\subsection{Test Data}
\label{sec:test-data}
Open and blind test sets were specifically created and curated for the development and test phases of the challenge, respectively. The blind test set was only available from the start of the 24-hour submission period of the test phase. Both sets are now publicly available in the test data repository \cite{lrac-test-data}.

\subsubsection{Open test set} 

A 1000-utterance clean English test set was chosen from the development and test split of the training corpora. This set was curated using stricter thresholds in terms of noise and reverberation than those applied during the training data curation. Subsequently, a subset of the speech data was augmented with either noise, RIRs, or both, to create synthetic noisy and reverberant data. All speech, noise, and RIR files were withheld from the training data. 
Participants were advised to exclude those reserved subsets from training, for which scripts were provided in the challenge's data preparation repository \cite{lrac2025data}.

The noise level and reverberation strength were adjusted for each track in accordance with the task specifications—light noise and mild reverb for Track 1, and a broader range of noise and reverb intensities for Track 2.

\subsubsection{Blind test set}
\label{subsec:blind-test-set}

A summary of the blind test set components and the tests for which they were used can be found in Table \ref{tab:evaluation}. The blind test consists of the following subsets:
\begin{itemize}
    \item \textit{Clean speech:} A set, used in both tracks, comprising 90 clean English utterances, each 4--8 seconds long. It contains speech samples from both adults and children and ensures a balanced representation of genders. 
    \item \textit{Simultaneous talkers:} A set of 20 files of English conversations between multiple speakers whose turns partially overlap. Used to evaluate the resilience of Track 1 systems to simultaneous takers.
    \item \textit{Real-world light noise \& reverb:} A set of 90 real-world English recordings containing perceptually light noise and/or mild reverberation. Used to evaluate Track 1 codecs' robustness under everyday conditions.
    \item \textit{Real-world reverb:} A set of 60 real-world English recordings with varying degrees of perceived reverberation strength (approx. 20\% weak, 40\% moderate, and 40\% strong) and light ambient noise. Used to test dereverberation performance of Track 2 systems.
    \item \textit{Real-world speech in noise:} A set of 100 real-world recordings of speech in various types of noise at different noise levels. The estimated SNR values approximately range from \textminus 5 to 25~dB (mean $\pm$ SD: 8.15 $\pm$ 6.5~dB). The perceived noise level was also taken into consideration during curation. This set was used for evaluating denoising performance of Track 2 systems.
    \item \textit{Speech intelligibility:} A set of 380 clean single-word English recordings \cite{lrac-test-data} selected from the multilingual speech intelligibility test set \cite{lechler_crowdsourced_2024} to form 384 test pairs. Used for both tracks to assess impact of codecs on intelligibility. Additionally, to evaluate the impact of denoising on intelligibility in Track 2, clean files were mixed with speech-shaped noise (SSN) at an SNR of 5~dB \cite{lrac-test-data}.
\end{itemize}
All test files were provided at the 24 kHz sampling rate. The majority of the real-world files were collected by Cisco Systems through crowdsourcing platforms such as Prolific and Amazon MTurk, representing natural noise, reverberation, and device variability. 
Note that the blind test set, used for the final evaluation, was completely disjoint from the training and open test set data pools.

\section{Baseline Systems}
\label{sec:baseline}

The challenge baseline models are open-source NACs implemented in a public fork of the ESPnet toolkit~\cite{lrac2025baseline}. Both Track 1 and Track 2 baselines employ convolutional encoder-decoder architectures with residual vector quantization and generative adversarial training. The Track 1 baseline was trained on clean speech, while the Track 2 baseline was trained on noisy and reverberant inputs (refer to \cite{lrac2025data}) with clean targets. Latency and computational complexity were tightly constrained to comply with the strict challenge requirements.
Model weights and a design spreadsheet detailing hyperparameters, latency, and complexity calculations were made available in via a public repository \cite{lrac2025baseline}. A public repository~\cite{lrac2025data} was also provided for training data downloading, preprocessing, and selection of the curated subset explained in Section~\ref{sec:training-data}.

For baseline systems for both tracks, the encoder processes raw audio using convolutional blocks consisting of residual convolution stacks followed by strided convolutions for temporal downsampling, producing embeddings every 10 ms. The decoder mirrors this structure, using transposed convolutions for upsampling followed by residual convolution sub-blocks. The residual vector quantizer comprises six layers, each containing 1,024 codewords contributing 1 kbps to the system bitrate. Input and output projection layers reduce computational cost. 

Models were trained for target bitrates of 1 kbps and 6 kbps, with random quantizer dropout. Codebooks were optimized via exponential moving average updates, while projection matrices were trained through backpropagation using straight-through gradient estimation to approximate gradients. Codeword selection was based on Euclidean distance. The training objective was a weighted sum of multi-scale mel-spectrogram loss~\cite{yamamoto-wavegan}, commitment loss~\cite{vq-vae}, adversarial loss from multi-scale feature discriminators~\cite{Defossez2022}, and feature matching loss on intermediate discriminator representations. A detailed description of the baseline systems is provided in our accompanying paper~\cite{lrac-baseline}.

\section{Evaluation}
\label{sec:evaluation}

The evaluation methodology was designed to assess both perceptual quality and intelligibility of submitted systems under controlled (synthetic) and realistic acoustic conditions. The evaluation was conducted across the development and test phases of the challenge. In the development phase, participants received continuous feedback via objective scores, and in the test phase, the final ranking was determined from large-scale crowdsourced listening tests. Objective metrics were \emph{not} used for the final ranking because they may exhibit limited reliability for models trained with generative methods \cite{lechler2025}.

\subsection{Development Phase}

During the development phase, objective metrics were reported through the challenge leaderboard to support iterative system improvement. We selected the following metrics, as they demonstrated strong correlations with listening tests for NACs under clean conditions \cite{mack2025assessing}: SCOREQ (non-matching reference mode) \cite{ragano2024scoreq}, UTMOS \cite{saeki2022utmos}, SHEET-SSQA \cite{huang2024sheetssqa}, and Audiobox Aesthetics Content Enjoyment (AE-CE) \cite{tjandra2025metaaudioboxaesthetics}. We also included PESQ \cite{rix_perceptual_2001}, as it is a widely adopted metric for assessing speech quality, despite showing weaker correlation with listening tests for NACs \cite{mack2025assessing}. The same set of metrics was utilized for the clean, noisy, and reverberant subsets of the open test set. The original clean signal served as the reference for intrusive metrics, namely SCOREQ and PESQ. 

\subsection{Test Phase}

During the test phase, the systems were evaluated using large-scale crowdsourcing listening tests. All systems submitted to the challenge underwent evaluation for clean speech quality and intelligibility (Secs.~\ref{sec:clean-speech-quality} and \ref{sec:intelligibility}, respectively). Additionally, Track 1 submissions were assessed for speech degradation in real-world scenarios (Sec.~\ref{sec:speech-degradation}) while Track 2 submissions were evaluated for speech enhancement performance under noisy and reverberant conditions (Sec.~\ref{sec:se-performance}). Table~\ref{tab:evaluation} provides a summary of the subjective evaluation battery. 
Intelligibility assessments were conducted using clean speech data for all models and speech in SSN for Track 2 models.

All listening tests were conducted on Prolific. Participants provided their informed consent prior to participation. The average hourly compensation was \$8.50, which exceeded Prolific's minimum payment rate. The participant pool was filtered to include only native English speakers without hearing difficulties or cochlear implants, with a minimum 98\% approval rate and at least 110 approved studies. We collected 15 responses per test item for DRT and 8 responses per file for other test types. Listener ratings were filtered via validation questions, attention checks, and hearing test thresholds. 

The raw test scores were obtained by first averaging ratings for individual test questions, and then averaging these to form an overall mean opinion score. Subsequently, the mean scores for each subset of the test were normalized to the range of $[0, 100]$, and the system's final score was calculated via a weighted sum of these normalized scores. The score ranges for the various tests, along with the weights for each bitrate mode and test condition are provided in Table \ref{tab:evaluation}. Further details regarding the normalization of raw test scores are available in the challenge rules \cite{lrac2025rules}.

\subsubsection{Clean Speech Quality}
\label{sec:clean-speech-quality}

Preserving speech quality is an essential aspect of a speech codec's performance. The quality of coded speech is often assessed with multiple stimuli with hidden reference and anchor (MUSHRA) tests \cite{itu_bs1534_2015}. While MUSHRA is a very precise test for comparing a few conditions, it is not practical to scale it for benchmarking a large number of conditions. Therefore, Absolute Category Rating (ACR) \cite{itu_p800_1996} is often preferred for benchmarking.
To balance test accuracy and scalability, we adopted the recently proposed MUSHRA--1S \cite{lechler2025mushra1s} test methodology to evaluate speech quality. Here, a single condition is tested at a time, alongside a reference and an anchor. This setup allows to freely scale up the number of test conditions while avoiding range-equalizing biases found in ACR tests \cite{cooper_investigating_2023} and context effects inherent to standard MUSHRA tests \cite{mushra_biases}. 
The ultralow bitrate (ULB) and low bitrate (LB) modes were evaluated using this assessment in both tracks (see 1a and 2a in Table \ref{tab:evaluation}). The same set of anchor (Opus at 6~kbps) and reference (original) signals was used throughout, enabling the reliable merging of multiple test results.

\subsubsection{Speech Degradation}
\label{sec:speech-degradation}

Speech codecs must be robust to a variety of acoustic conditions to ensure reliable performance in real-world scenarios. While transparency codecs are not expected to enhance or mitigate adverse acoustic conditions, they should not introduce speech degradations in those settings. We tested Track 1 systems (Table \ref{tab:evaluation}, 1b and 1c) for speech in light noise and reverberation, and overlapping speech. We utilized degradation category rating (DCR) tests \cite{itu_p800_1996},  instructing listeners to penalize only degradations in speech compared to the real-world reference. 
This approach focuses on penalizing degradation of speech and prevents raters from favoring solutions that denoise or dereverberate.

\subsubsection{Speech Enhancement Performance}
\label{sec:se-performance}

We evaluated the speech denoising and dereverberation performance of Track 2 systems (Table \ref{tab:evaluation}, 2b and 2c) on real-world data through ACR tests. ACR is commonly used \cite{dns2023,urgent} for speech enhancement evaluation, as it is a simple and fast approach that can evaluate a large number of test conditions and does not require pristine references, which are typically unavailable for real-world recordings.

\subsubsection{Intelligibility}
\label{sec:intelligibility}

It is essential to assess the effect of speech processing systems on intelligibility of speech. While various test methods for exist for this task, we selected the diagnostic rhyme test (DRT) \cite{voiers_evaluation_1965} which has been shown to be repeatable and strongly correlated with other speech intelligibility methods \cite{kryter_comparisons_1965} while providing acoustic and linguistic insights and having shorter test times. We follow a recent study which adapts DRT to crowdsourcing platforms \cite{lechler_crowdsourced_2024}. 
DRT tests were conducted at the ULB for clean conditions in both tracks (Table \ref{tab:evaluation}, 1d and 2d) to avoid a potential ceiling effect for the LB setting.
Additionally, Track 2 systems were evaluated on speech in noise to assess the impact of enhancement on intelligibility (Table \ref{tab:evaluation}, 2e). 

\section{Conclusion}

In this work, we have presented the motivation, scope, and design considerations for the inaugural Low-Resource Audio Codec Challenge. During the test phase, we received six unique submissions for Track 1 and nine for Track 2, excluding baseline entries. The final ranking based on crowdsourced listening tests, as well as the system description reports, were made publicly available on the challenge website. A comprehensive analysis of the submitted systems and their performance will be addressed in future work.

\section{Acknowledgments}
\label{sec:typestyle}

The authors thank the Cisco Collaboration AI Data Team---Daniel Arismendi, Ayoub Zaidour, James Taylor, Alexandra Meier---for dataset curation and augmentation support, and the Crowdsourcing Team---Tarek Afifi, Miguel Plaza Rosillon, Ana Rivera Jaramillo---for challenge website and crowdsourcing infrastructure assistance.

\bibliographystyle{paper}
\bibliography{paper}

\end{document}